\documentclass[11pt,twoside]{article}  
\usepackage{adassconf}

\begin{document}   

%
%

\paperID{P.156}

%
%
%
%

\title{Autonomous Observing and Control Systems for PAIRITEL, a 1.3m
Infrared Imaging Telescope}

\titlemark{PAIRITEL Automated Observing, Control Systems}

%

\author{J. S.\ Bloom\altaffilmark{1}, Dan L.\
  Starr\altaffilmark{2}, Cullen H.\ Blake\altaffilmark{3}, M. F. Skrutskie\altaffilmark{4}, Emilio E.\ Falco\altaffilmark{3}}
\altaffiltext{1}{University of California, Berkeley, CA, USA}
\altaffiltext{2}{Gemini North Observatory, HI, USA}
\altaffiltext{3}{CfA/Harvard, MA, USA}
\altaffiltext{4}{University of Virginia, VA, USA}


\contact{Joshua S. Bloom}
\email{jbloom@astro.berkeley.edu}

%
%
%
%
%

\paindex{Bloom, J. S.}
\aindex{Starr, D.}     
\aindex{Blake, C.}     
\aindex{Falco, E.}     
\aindex{Skrutskie, M.}     

%
%

\authormark{Bloom, Starr, Blake \& Falco}




\begin{abstract}          

The Peters Automated Infrared Imaging Telescope (PAIRITEL) is the
first meter-class telescope operating as a fully robotic IR imaging
system. Dedicated in October 2004, PAIRITEL began regular observations
in mid-December 2004 as part of a 1.5 year commissioning period. The
system was designed to respond without human intervention to new
gamma-ray burst transients: this milestone was finally reached on
November 9, 2005 but the telescope had a number of semi-automated
sub-10 minute responses throughout early commissioning.  When not
operating in Target of Opportunity mode, PAIRITEL performs a number of
queue scheduled transient monitoring campaigns. To achieve this level
of automation, we have developed communicating tools to connect the
various sub-systems: an intelligent queue scheduling database,
run-time configurable observation sequence software, a data reduction
pipeline, and a master state machine which monitors and controls all
functions within and affecting the observatory.

\end{abstract}


\section{Project background and constraints}

PAIRITEL, name for the late telescope operator Jim Peters, 
is an automated 1.3m telescope located at the ridge of Mt.\ Hopkins in
Arizona.  The telescope and simultaneous $J$,$H$,$K_s$ camera were
formerly used in the 2MASS project which ended data taking operations
in 2001. Our refurbishing work began in mid-2003 using a small
donation from the Harvard Milton Fund. Basic telescope automation was
achieved by the dedication on October 21st, 2004.

From the outset, PAIRITEL was designed for automated, queue based
observing, with the ability to rapidly respond to targets of
opportunity (ToOs) such as gamma-ray burst (GRB) alerts from
space-based satellites.  The best response thus far was 90 seconds
from gamma-ray burst trigger (GRB 051109a) to the beginning of the
first observation. Although rapid response telescopes require a
significant amount of automation, the remote location of this
telescope required additional levels of autonomy.

We required that the observatory software be able to diagnose
inclement weather and control system problems and respond
appropriately, as if an observer were present.  PAIRITEL determines
the observing schedule using an intelligent queuing database which
incorporates new observations into a dynamic scheduling system, while
accounting for recently acquired observations.  Software components
must also rely upon cross communication.  For example, a monitor of
the transmission, acquisition/reduction software, and telescope
pointing need complex interactions to maximize observing efficiency.

\section{Master daemon}

The key to coordinating our autonomous observatory is a program called
the `master daemon' (hereafter {\sc masterd}).  Based on a robotics
state machine, {\sc masterd} monitors the states of all other software
processes.  Then, using a template which defines actions for various
state changes, it commands other software atoms to act.

Separating each observatory task into individual software
sub-processes (`daemons') allows them to be managed in a generalized
way (Fig.\ 1).  As daemons are initiated and restartable on a system
level, we can then assume them always to be running.  If a critical
error occurs in a daemon, the master daemon will notice the timeout of
a `heartbeat', and restart it.

\begin{figure}
\epsscale{.85}
\plotone{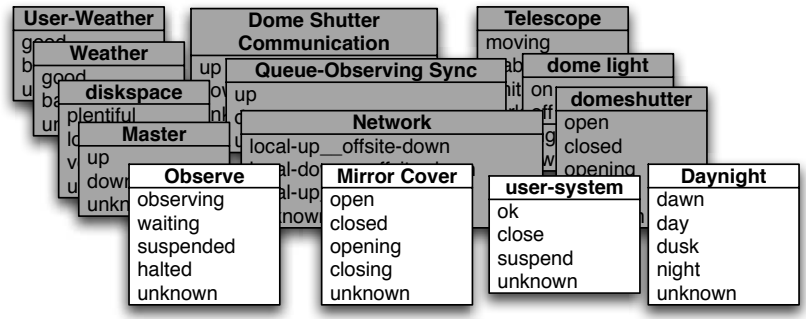}
\caption{Sub-systems, each with their own daemon control, whose states are monitored by 
the `master daemon'. Shown are some example sub-systems with corresponding sets of states. Since
these states are enumerated, {\sc masterd} contends with a finite
number of system states. Matching of state changes and actions taken
during state change are prescribed in a list of rules. The logic for
determining the value of the sub-system states are encoded in the
sub-system daemons.}
  \label{P.156-fig-1}
\end{figure}

The state machine design also allows for self-regulating actions.  One
example is determining the current accuracy of the telescope pointing.
By monitoring a transmission daemon (which determines sky
transparency) and possibly the time of the night, the master daemon
can see when a pointing check/correction is needed.  It then commands
other daemons to do actions such as: halting an observation,
performing a ``soft" pointing check at a known bright star position, or
initiating a hardware-based pointing check by using a custom set of
codes designed to interact with opto-interruptors and a tiltometer.  In
this case, once the pointing state is "excellent", the master daemon
will resume the observations.


\section{Observation software}

The observation software consists of three main parts: the observation
database, the scheduling software, and the observing daemon.  Although
this software is controlled by the master daemon on the highest level,
the queuing, acquisition, and time accounting takes place independent
of it. The observation database is MySQL based and generally accessed
remotely by astronomers using a PHP interface.  The scheduling
software accesses the database using the MySQLdb Python module.  This
database is organized as tables hierarchically by: project, then
objects within projects, and observations of objects.  Users can set
priorities of objects within their projects; the projects themselves
are assigned relative priorities and total awarded observing
time. Bookkeeping for total time observed per project is performed in
a realtime feedback system.

The observation queue scheduler selects which programs and objects are
observed during a night (Fig.\ 1).  It calculates the observing plans
on a daily basis, but can also be commanded to recalculate them
realtime by an astronomer or by itself (for instance when a ToO
arrives).  The resulting queues are optimized using variables such as
airmass, priority, and time since previous observation.  The scheduler
also dynamically updates its database by parsing recently acquired
FITS files and adding the information to the corresponding programs.
\begin{figure}
\epsscale{.70}
\plotone{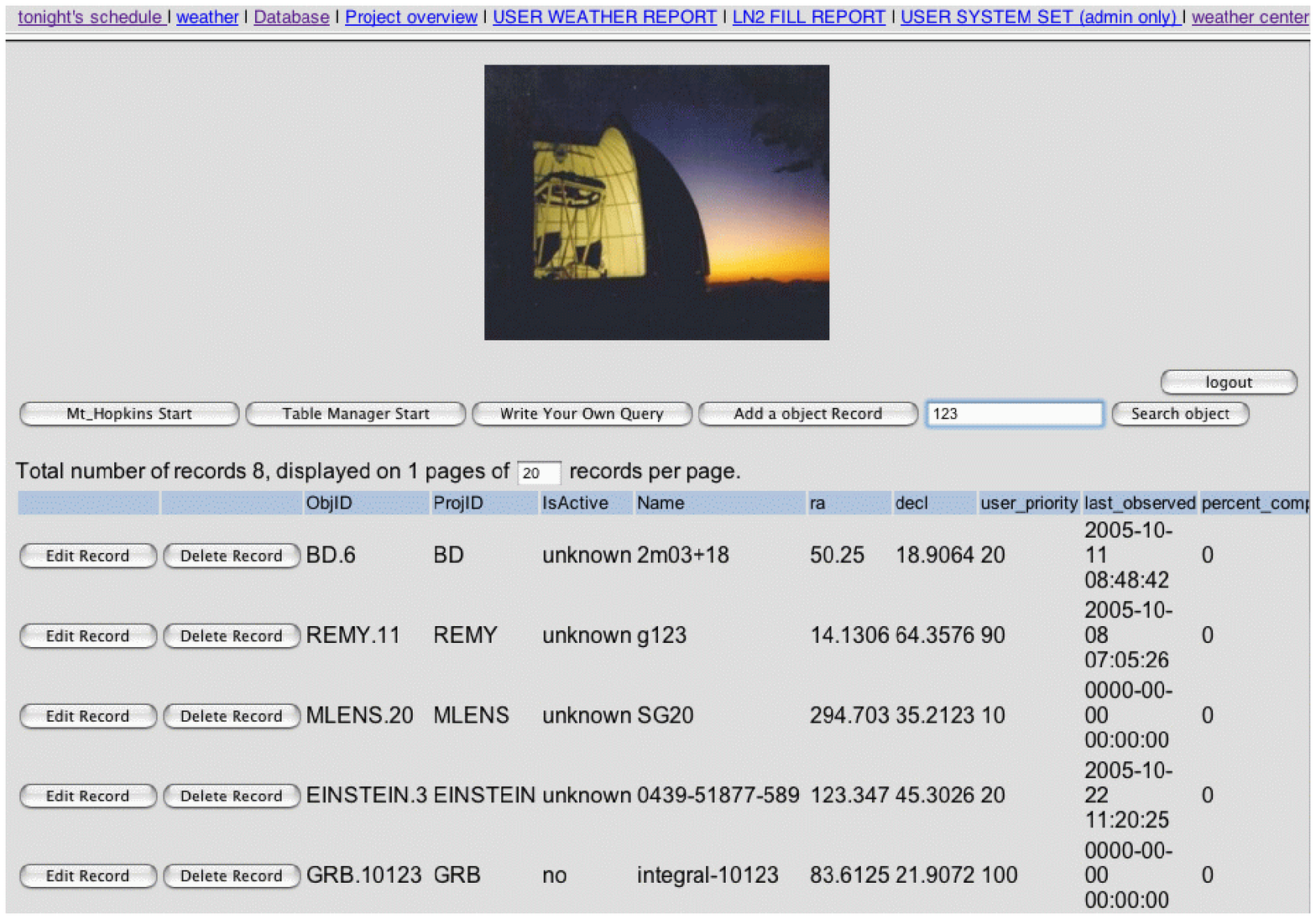}
\caption{Snapshot of the database interface, a PHP environment built around the PAIRITEL MySQL database.}
  \label{P.156-fig-4}
\end{figure}
The observing daemon executes observation sequences sent by the
scheduling software.  It calculates all dithering sequences and using
python interfaces, it controls both the telescope and camera controllers.

The requested observation sequences (and the output metadata from the camera)
are defined in XML files and are automatically queued by the observing daemon
whenever the scheduler sends them.  In the case that the observation is a ToO,
the daemon aborts the current exposure, slews the telescope, and begins
exposing. The PAIRITEL reduction pipeline is written in Python using the {\it
pyfits} and pyraf modules.  The image processing is fairly straight forward,
using realtime created darks and archived twilight flats.  In addition, bad
pixels are fixed and the world coordinate WCS is calculated in the final
mosaiced image.
\begin{figure}
\epsscale{.70}
\plotone{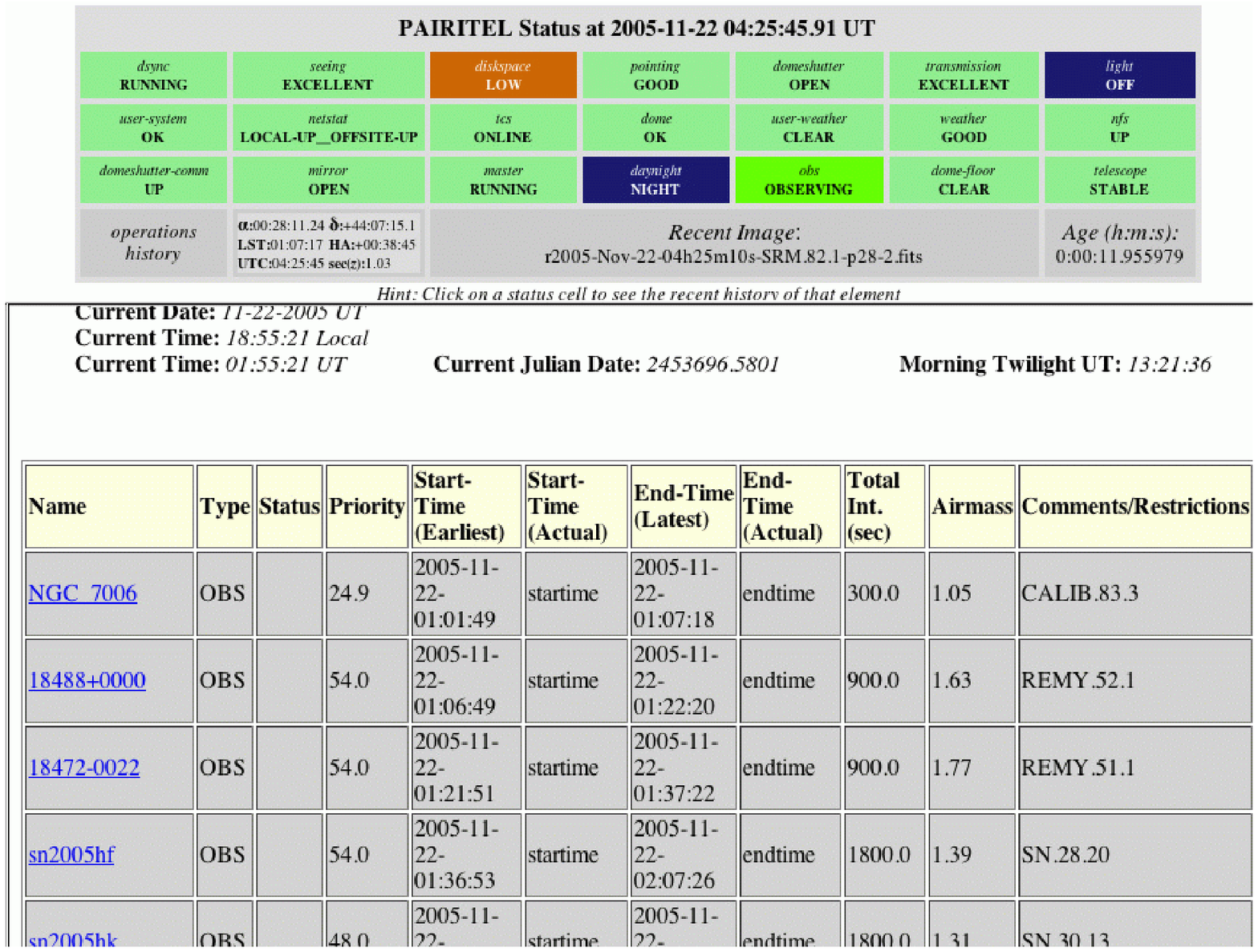}
\caption{Snapshot of the master daemon status web page and views of the current schedule. See {\tt http://status.pairitel.org/}}
  \label{P.156-fig-5}
\end{figure}
The `status' web page (Fig.\ 3) displays an overview of all states
monitored by the master daemon.  The color coded boxes allow a quick
assessment of the system.  Additionally, clicking on any state gives a
detailed history log of that state.  Real-time quick-look reductions are
also available with a click.

\section{Summary}

Automating the PAIRITEL telescope has been surprisingly fast due to
quick software implementation.  We attribute much of this to our
choice of Python as our development language.  Python has allowed easy
coding of software hooks into electronic devices using serial,
network, and parallel port modules.  The high level aspects of the
language also helped us design algorithms which are template based and
thus easily configured and updateable.  The overall ease in producing
working code has allowed us to experiment with different approaches in
solving problems.  For these reasons, we were able to rapidly develop
and test the master deamon and observation software.

For more information, visit: {\tt http://www.pairitel.org/}.


\end{document}